\documentstyle[epsfig]{aprim10}
\input{epsf}

\newif\ifAMStwofonts
\AMStwofontstrue

\ifoldfss
  \ifCUPmtlplainloaded \else
    \NewTextAlphabet{textbfit} {cmbxti10} {}
    \NewTextAlphabet{textbfss} {cmssbx10} {}
    \NewMathAlphabet{mathbfit} {cmbxti10} {} 
    \NewMathAlphabet{mathbfss} {cmssbx10} {} 
  \fi
  \ifAMStwofonts
    \ifCUPmtlplainloaded \else
      \NewSymbolFont{upmath} {eurm10}
      \NewSymbolFont{AMSa} {msam10}
      \NewMathSymbol{\upi}     {0}{upmath}{19}
      \NewMathSymbol{\umu}     {0}{upmath}{16}
      \NewMathSymbol{\upartial}{0}{upmath}{40}
      \NewMathSymbol{\leqslant}{3}{AMSa}{36}
      \NewMathSymbol{\geqslant}{3}{AMSa}{3E}

    \fi
  \fi
\fi 

\ifnfssone
  \newmathalphabet{\mathit}
  \addtoversion{normal}{\mathit}{cmr}{m}{it}
  \addtoversion{bold}{\mathit}{cmr}{bx}{it}
  \newmathalphabet{\mathbfit} 
  \addtoversion{normal}{\mathbfit}{cmr}{bx}{it}
  \addtoversion{bold}{\mathbfit}{cmr}{bx}{it}
  \newmathalphabet{\mathbfss} 
  \addtoversion{normal}{\mathbfss}{cmss}{bx}{n}
  \addtoversion{bold}{\mathbfss}{cmss}{bx}{n}
  \ifAMStwofonts
    \ifCUPmtlplainloaded \else
      \UseAMStwoboldmath
      \makeatletter
      \new@mathgroup\upmath@group
      \define@mathgroup\mv@normal\upmath@group{eur}{m}{n}
      \define@mathgroup\mv@bold\upmath@group{eur}{b}{n}
      \edef\UPM{\hexnumber\upmath@group}
      \new@mathgroup\amsa@group
      \define@mathgroup\mv@normal\amsa@group{msa}{m}{n}
      \define@mathgroup\mv@bold\amsa@group{msa}{m}{n}
      \edef\AMSa{\hexnumber\amsa@group}
      \makeatother
      \mathchardef\upi="0\UPM19
      \mathchardef\umu="0\UPM16
      \mathchardef\upartial="0\UPM40
      \mathchardef\leqslant="3\AMSa36
      \mathchardef\geqslant="3\AMSa3E
    \fi
  \fi
\fi 

\ifnfsstwo
  \DeclareMathAlphabet{\mathbfit}{OT1}{cmr}{bx}{it}
  \SetMathAlphabet\mathbfit{bold}{OT1}{cmr}{bx}{it}
  \DeclareMathAlphabet{\mathbfss}{OT1}{cmss}{bx}{n}
  \SetMathAlphabet\mathbfss{bold}{OT1}{cmss}{bx}{n}
  \ifAMStwofonts
    \ifCUPmtlplainloaded \else
      \DeclareSymbolFont{UPM}{U}{eur}{m}{n}
      \SetSymbolFont{UPM}{bold}{U}{eur}{b}{n}
      \DeclareSymbolFont{AMSa}{U}{msa}{m}{n}
      \DeclareMathSymbol{\upi}{0}{UPM}{"19}
      \DeclareMathSymbol{\umu}{0}{UPM}{"16}
      \DeclareMathSymbol{\upartial}{0}{UPM}{"40}
      \DeclareMathSymbol{\leqslant}{3}{AMSa}{"36}
      \DeclareMathSymbol{\geqslant}{3}{AMSa}{"3E}
    \fi
  \fi
\fi 

\ifCUPmtlplainloaded \else
  \ifAMStwofonts \else 
    \def\upi{\pi}
    \def\umu{\mu}
    \def\upartial{\partial}
  \fi
\fi

\title[Force-Free Pulsar Light Curves]{Modeling Pulsar Gamma-Ray Light
Curves Using Realistic Magnetospheric Geometries}

\author[Bai \& Spitkovsky]
       {X.-N. Bai$^1$ and A. Spitkovsky$^1$\\
        $^1$Department of Astrophysical Sciences, Princeton University, Princeton, NJ, 08544, USA}
\date{}

\pagerange{\pageref{firstpage}--\pageref{lastpage}}
\pubyear{2008}

\begin{document}

\maketitle

\label{firstpage}

\begin{abstract}
Gamma-ray emission from pulsars is thought to arise from
accelerating regions in pulsar's outer magnetosphere. The shape of
the light curves is thus sensitive to the details of the magnetic
geometry of the magnetosphere. In this work, we show the first
calculations of light curves from the more realistic force-free
field under the framework of conventional emission models. We
compare the properties of gamma-ray emission between the commonly
used vacuum dipole magnetic field and the new force-free field.
We discuss the role of the polar cap shape and aberration effect
on the appearance of the light curves as well as the formation of
caustics on the sky map. With the force-free field, the double-peak
pulse profile is best reproduced if the emission zone lies in a thin
layer just outside the current sheet, and the peaks are mainly
contributed from regions near the light cylinder. The conventional
two-pole caustic model can produce up to four peaks in general, while
the conventional outer-gap model can normally produce only one peak.
These results will be useful for interpreting Fermi telescope observations.
\end{abstract}


\section{Introduction}

Pulsars are rotating neutron stars (NS) with strong magnetic field.
Seven of them were detected as gamma-ray pulsars with EGRET
\cite{th04}. The light curves of these gamma-ray pulsars are
typically double-peaked, and have substantial off-peak emission.
Theoretical models have been developed in order to explain the
nature of pulsar gamma-ray emission, namely, the polar-cap (PC)
model \cite{dh82,dh96}, the slot-gap (SG, or two-pole caustic, TPC
for short) model \cite{arons79,arons83,dr03,dhr04}, and the
outer-gap (OG) model \cite{chr86,ry95,crz00}. In these models,
particles are accelerated to ultra-relativistic energies in ``gap"
regions where strong electric fields are developed due to the
deficit of charge. The gamma ray emission comes from curvature and
inverse Compton (IC) radiation of these energetic particles. Based
on vacuum magnetic field geometries, SG and OG models are more
favored since they can reasonably well reproduce the double-peak
light curves.

However, pulsar magnetosphere is filled with plasma \cite{gj69}. The
plasma is essentially force-free (FF), with $\rho{\bf E}+{\bf
j}\times{\bf B}/c=0$, where $\rho$ and ${\bf j}$ are charge and
current density. In the presence of plasma, the magnetosphere
consists of open and closed field line regions. Poloidal current
flows out along open field lines and induces toroidal magnetic
field, which dominates beyond the light cylinder (LC for short,
$R_{\rm LC}=c/\Omega$). Recently, Spitkovsky \shortcite{as06}
obtained the full 3D magnetospheric structure using FF simulations.
The FF field geometry differs substantially from the vacuum field
geometry (i.e., the retarded dipole field), which is commonly used
in light curve calculations, especially around the LC. The light
curve is very sensitive to the geometry of the emission zones, thus
to the field geometry itself. Therefore, it is important to revisit
pulsar high-energy emission models using the more realistic FF field
geometry.


This paper is organized as follows. In section 2 we show the light
curves from the vacuum field geometry, pointing out that previously
calculated vacuum light curves should be modified when aberration
effect is correctly treated. In section 3 we present the light
curves from the FF field using the two-pole caustic model, and
mention the applications to the outer-gap model.

\section{Relativistic Effects and Vacuum Field Result Corrections}

In order to calculate the light curve, we need to find the emitting
region in the magnetosphere. In the TPC model, the emission zone is
assumed to be along the last open field lines (LOFLs) starting from
the polar cap. In the OG model, the radiation region is assumed to
be beyond the null-charge surface and along open field lines.
Technically, given a magnetic field geometry (either vacuum or FF
field), field lines are considered open if they can cross the LC in
the lab (observer's) frame (LF). We trace magnetic field lines in
the LF and find the LOFLs. The polar cap is the region on the NS
surface where open field lines originate. We calculate the magnetic
colatitude of polar cap rim $\theta_m^{\rm{rim}}$ at fixed magnetic
azimuth $\phi_m$ (the subscript ``$m$" means w.r.t. magnetic axis
rather than rotation axis), and define the open volume coordinate as
$r_{\rm ov}=\theta_m /\theta_m^{\rm rim}$. In the TPC model
\cite{dr03,dhr04} emission comes from a thin layer centered at LOFL
($r_{\rm ov}=1$) extended from the NS surface up to a certain radius
($R\sim0.75R_{\rm LC}, r\sim1.0R_{\rm LC}$ where $r$ is distance to
the NS center and $R$ is cylindrical radius). In the OG model
\cite{crz00} the emission comes from a thicker layer centered at
$r_{\rm ov}\sim0.90$ extended from the null charge surface (where
$B_z=0$) to the LC.

For curvature/IC radiation, the radiation direction should be along
the direction of particle motion. An ultra-relativistic particle
moves at the speed of light, and its velocity can be decomposed into
corotation velocity (i.e., drift velocity) plus a component along
the direction of the magnetic field. Thus, the radiation direction
$\vec{\eta}$ in the LF is determined by
\begin{equation}
\vec{\eta}=f{\bf B}+\vec{\beta}\ ,\label{eq:1}
\end{equation}
where $\vec{\beta}=\vec{\Omega}\times{\bf r}/c$ is the normalized
corotation velocity, ${\bf B}$ is the magnetic field in the LF, and
$f$ is a coefficient determined by the requirement that
$|\vec{\eta}|=1$.

This approach is different from earlier works on OG and TPC models
with vacuum field geometries [e.g. Romani \& Yadigaroglu
\shortcite{ry95}, Cheng et al. \shortcite{crz00}, Dyks et
al.\shortcite{dhr04}]. In previous works, it was implicitly assumed
that the vacuum field, which is the solution of vacuum Maxwell
equations in the LF, is valid in the instantaneously corotating frame
(ICF). In other words, in eqn.(\ref{eq:1}), ${\bf B}$ was improperly replaced by ${\bf B}'$,
where ${\bf B}'$ is obtained by the Lorentz transformation of ${\bf
B}$ from ICF to LF [see Bai \& Spitkovsky \shortcite{bs08a} for a
more detailed discussion]. Therefore, results from these models
require revision.

Photons emitted from the emission zones are collected in the sky map
($\phi,\xi_{\rm obs}$), where $\phi$ is the phase of rotation with
corrections for photon travel time (i.e., time delay effect), and
$\xi_{\rm obs}$ is observer's viewing angle. We assume constant
emissivity along field line. The light curve seen by the observer is
then obtained by cutting the sky map at a specified viewing angle
$\xi_{\rm obs}$.

\begin{figure} 
 \centerline{\epsfxsize=9cm\epsffile{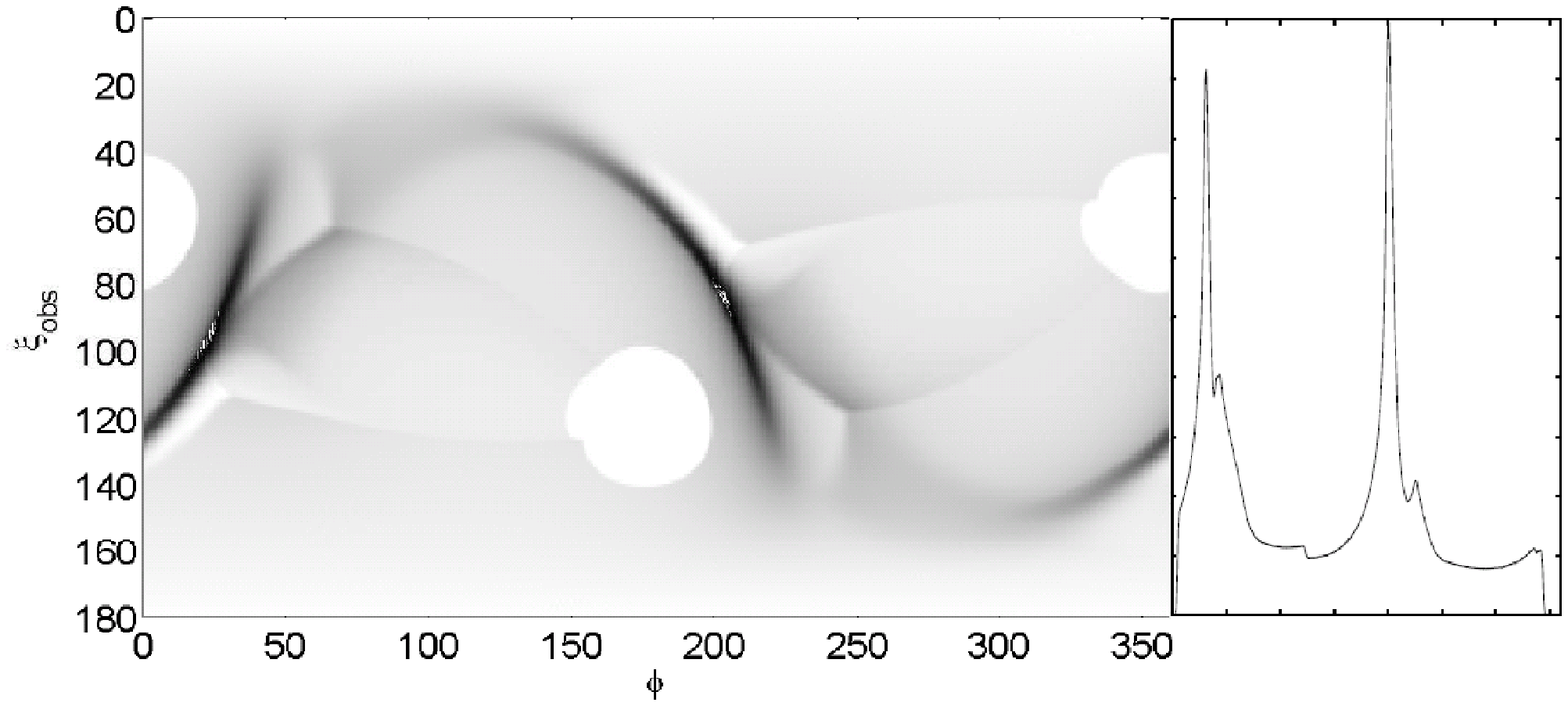}}
 \centerline{\epsfxsize=9cm\epsffile{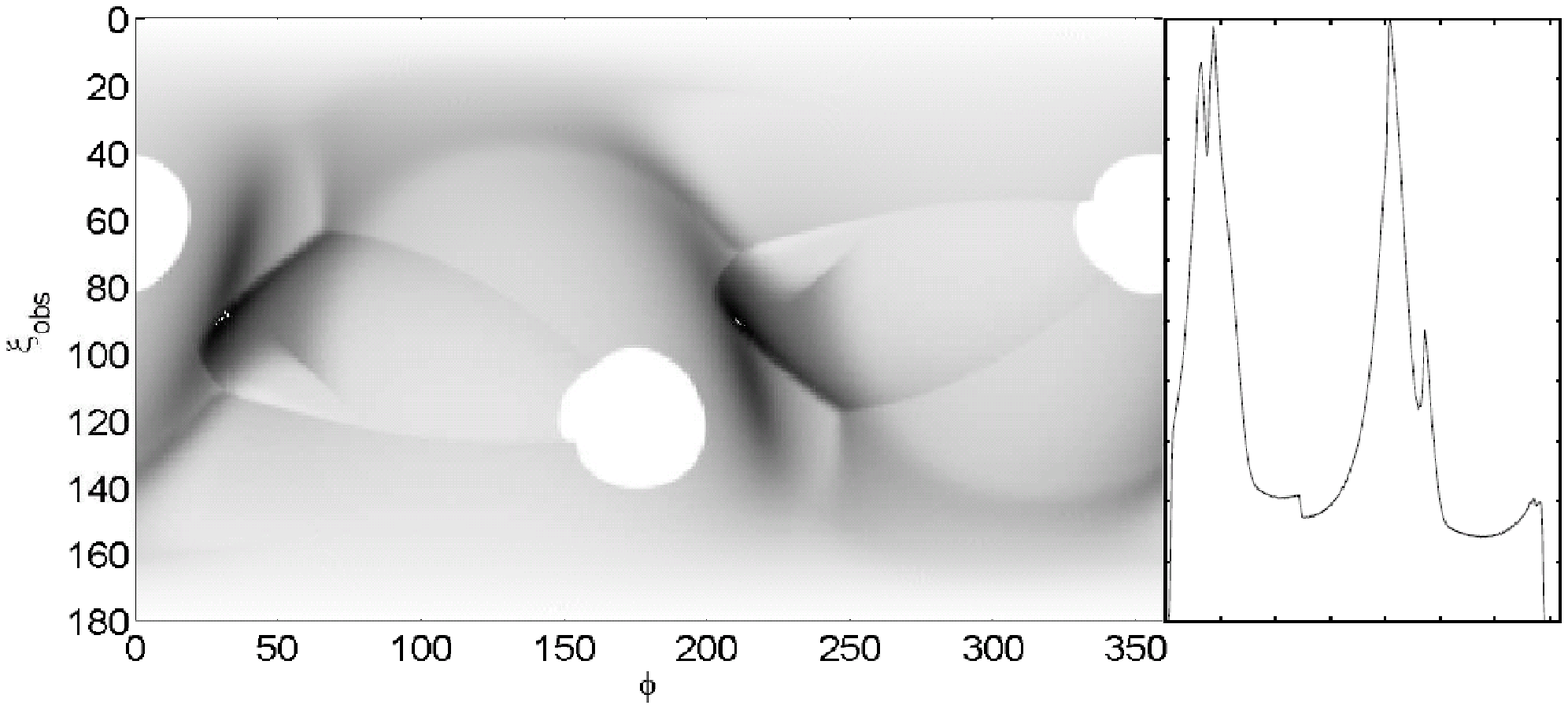}}
 \caption[]{Sky map (left) and light curve (right) from two-pole caustic
 model using vacuum field. Inclination angle $\alpha=60^{\circ}$,
 viewing angle $\xi_{\rm obs}=80^{\circ}$. Up: with inappropriate
 treatment of aberration; down: with the correct aberration
 [eqn.(\ref{eq:1})].
formula.}
\end{figure}

In Fig.1 we plot the sky map and light curve from TPC model for an
oblique rotator with the inclination angle $\alpha=60^{\circ}$ using
vacuum field geometry. The upper plot is a reproduction of Dyks et
al. \shortcite{dhr04}'s result, where the retarded vacuum field was
implicitly assumed to be in the ICF. There are two strong caustics
associated with the two poles. They are formed at modest distance
from the NS (roughly $0.3-0.6R_{\rm LC}$). In the light curve, two
sharp peaks are thus produced. In the bottom plot we correct the
aberration with eqn.(\ref{eq:1}). We find that the caustics in the
upper plot are now much widened and weakened. There are still two
peaks, but much wider, and a substantial contribution to the peaks
comes from the overlap of emission from both poles, rather than from
the caustics. This result also holds for other inclination angles.
We conclude that the TPC model has difficulties in producing sharp
peaks with corrected aberration. We also note that the correction of
abberation formula also affect the conventional outer-gap models using
the vacuum field \cite{bs08a}.

\section{Results from Force-Free Field Geometry}

The advent of FF field \cite{as06} makes it possible to test
theoretical models with more realistic magnetospheric geometries.
The FF magnetosphere has ${\bf E}\cdot{\bf B}=0$ everywhere and thus
has no intrinsic particle acceleration. As long as pulsar's emission
power is much smaller than the total spin down energy loss rate
(which is almost always the case), the FF field provides a reliable
field structure in the magnetosphere. The pulsar's radiation comes
from a small region of magnetosphere (i.e., the emission zone) where
non-ideal deviations from the FF condition makes particle
acceleration possible. The current sheet is a key feature of FF
field. Beyond the LC, it separates field lines from the two poles
\cite{ckf99,bs08b}, and it is also a likely place for dissipation
and resistive heating \cite{ag07}. Consequently, the pulsar's
gamma-ray emission may be connected to the current sheet.

The sky map is determined by the field geometry as well as the shape
of the polar cap. Although the FF field appears similar to the
vacuum field near the NS surface, the shape of the polar cap is
different, because the polar cap shape is sensitive to the field
structure near the LC. The polar cap shape in the FF field is more
circular and larger than in the vacuum field \cite{bs08b}, and
it will lead to fundamental differences in the sky maps, as we will
illustrate here.

In Fig.2 we show the sky map of the conventional TPC model using the
FF field which is calculated with $r_{\rm ov}=1.0$. The emission
zones are assumed to be extended from the polar cap to the LC along
current sheet. Depending on observer's viewing angle, the sky map
can produce light curves with up to four widely separated peaks,
which are inconsistent with the current observations. We
have experimented with a number of other possible solutions, and
find that the double-peaked pulse profile can be reproduced if we
allow the emission zones to be located in a layer just outside the
current sheet. In Fig.3 (upper plot) we show the sky map and light
curve from the emission zone $r_{\rm ov}=0.90$ under this ``annular
gap" model (or modified TPC model). There are now two strong caustics
associated with both poles. The caustics here are formed around the LC.
The double-peak pattern is very robust as we vary the inclination angle
$\alpha$.

\begin{figure} 
 \centerline{\epsfxsize=7.5cm\epsffile{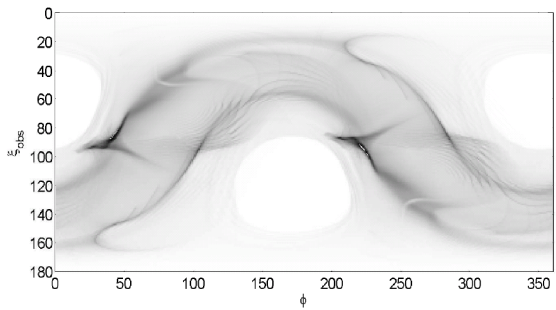}}
 \caption[]{Sky map from conventional two-pole caustic model $r_{\rm ov}=1.0$
 using force-free field. Inclination angle $\alpha=60^{\circ}$, viewing angle
 $\xi_{\rm obs}=80^{\circ}$.}
\end{figure}

\begin{figure} 
 \centerline{\epsfxsize=9cm\epsffile{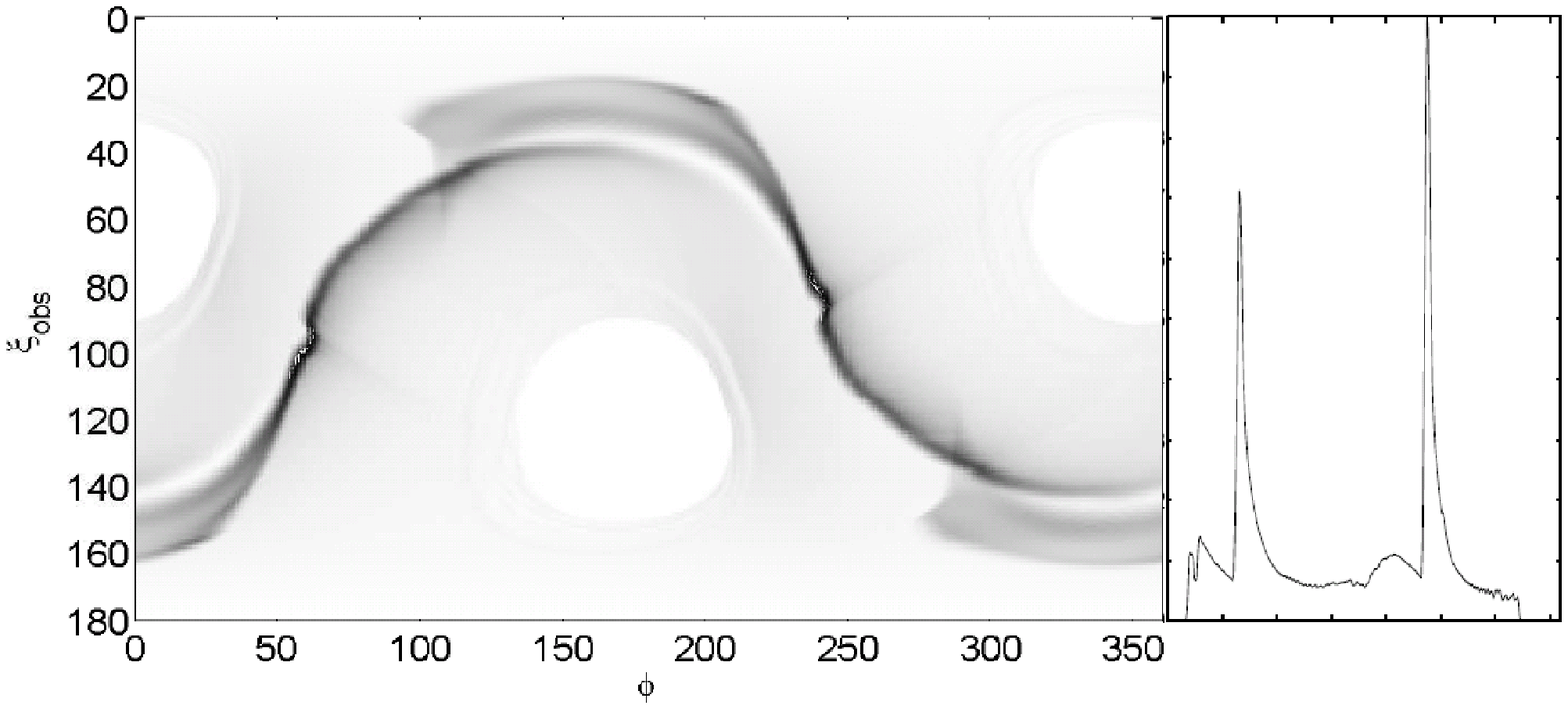}}
 \centerline{\epsfxsize=9cm\epsffile{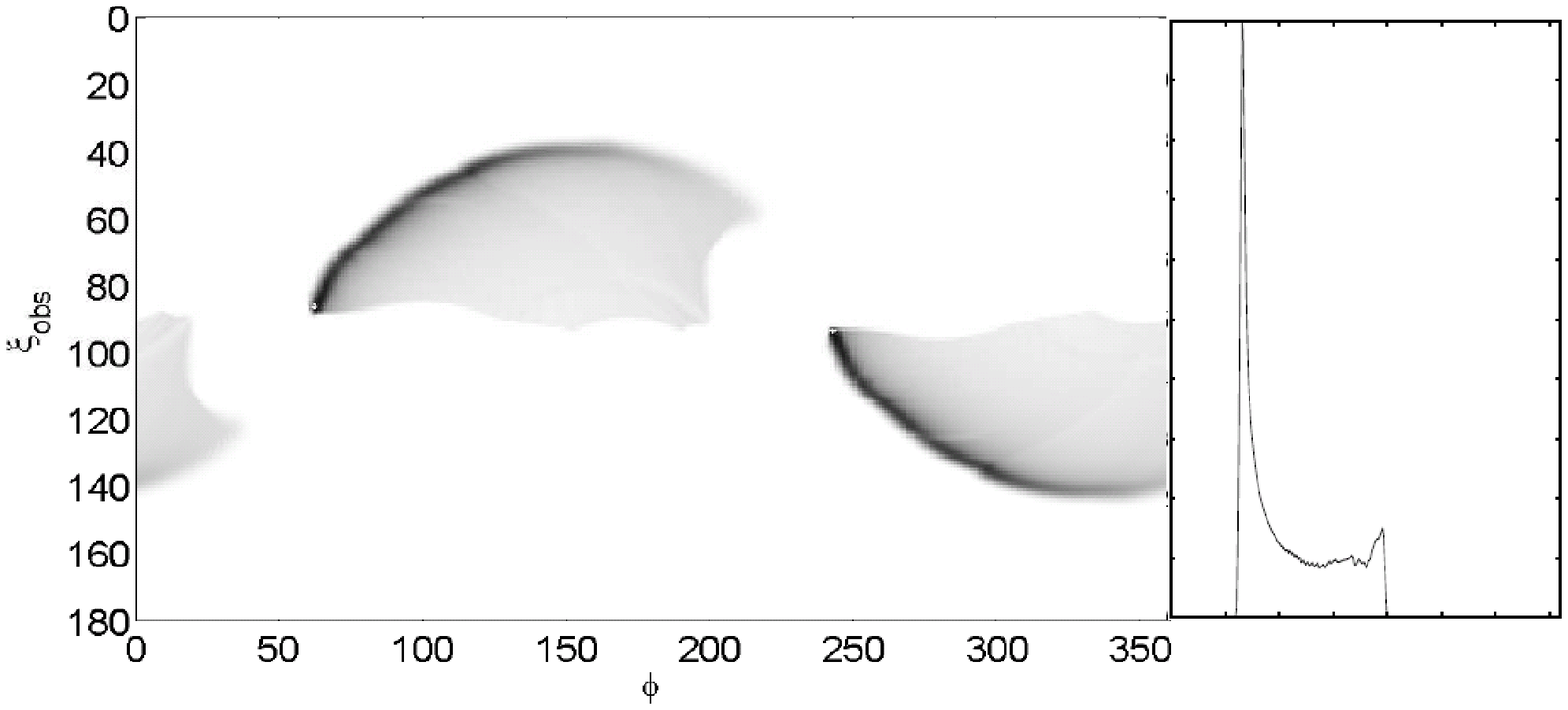}}
 \caption[]{Sky map (left) and light curve (right) from the annular gap model
 (a modified two-pole caustic model) with $r_{\rm ov}=0.90$ (up) and the
 conventional outer-gap model with $r_{\rm ov}=0.90$ (bottom) using force-free
 field. Inclination angle $\alpha=60^{\circ}$, viewing angle $\xi_{\rm obs}
 =80^{\circ}$.}
\end{figure}

The choice of the open volume coordinate $r_{\rm ov}=0.90$ is
similar to that used in the OG models. However, we emphasize here
that this is not an OG model. For an OG model, only field lines that
cross the null-charge surface can contribute \cite{tcs08}. Here we
have considered emissions from all field lines with a certain
$r_{\rm ov}$.  It turns out that at $\xi_{\rm obs}=60^{\circ}$, only
a small fraction of the field lines can cross the null-charge
surface, and only one peak survives (see bottom plot of Fig.3).

\section{Conclusion}

Existing models for pulsar's gamma-ray emission (i.e., slot-gap and
outer-gap models) are all based on the vacuum field geometries,
which carry large uncertainties. Hence, the light curves predicted
from these models are questionable. In this paper we revisit
pulsar's gamma-ray emission using the more realistic force-free
field configurations.

We first point out the inconsistent treatment of the aberration
effect in previous works and compare them with aberration corrected
results. The new results indicate that sharp peaks are difficult to
be produced using vacuum field geometry in the conventional two-pole
caustic model.

Using the force-free field, we find that the conventional slot-gap
(two-pole caustic) model tends to produce four peaks, while the
conventional outer-gap model is able to provide just one peak,
inconsistent with current observations. An ``annular gap" model, or
modified two-pole caustic model, where emissions are assumed to
originate from a thin layer just outside the entire current sheet,
is able to reproduce the basic observed features of gamma-ray
pulsars. The physics behind remains to be understood.

The launch of the Fermi Gamma-ray Space Telescope will significantly
expand the sample of gamma-ray pulsars and will provide very
accurate pulse profiles. Our results will be useful for the
interpretation of Fermi telescope observations and for probing the
physics of pulsar magnetospheres.

\section*{Acknowledgment}
This work is supported by NASA grant NNX08AW57G.

\label{lastpage}

\clearpage

\end{document}